\documentstyle[11pt,preprint, epsfig,graphicx,aps]{revtex}
\newcommand {\bi} {\bibitem}
\newcommand{\Cred} {}
\newcommand{\Cblu} {}
\newcommand{\via} {}
\newcommand {\bea} {\begin{eqnarray} }
\newcommand {\eea} {\nonumber \end{eqnarray}}
\newcommand {\be} {\begin{equation}}
\newcommand {\ee} {\end{equation}}
\newcommand {\eps} {\epsilon}
 \newcommand {\si} {\sigma}

\newcommand {\lan} {\langle}
\newcommand {\ran} {\rangle}

\newcommand {\cB}  {{\cal B}}
\newcommand {\cC}  {{\cal C}}

\newcommand {\cP}  {{\cal P}}

\newcommand {\bc} {\begin{center}}
\newcommand {\ec} {\end{center}}
\newcommand {\bd}{\begin{displaymath}}
\newcommand {\ed}{\end{displaymath}}

\newcommand {\tgh} {\mbox{th}}
\newcommand {\arth} {\mbox{arth}}

\newcommand {\for} {\ \ \ \mbox{for}\ \ }

\def \form#1 {eq. (\ref{#1}) }
\def \parziale#1#2  {{\partial {#1} \over \partial {#2}}}
\def \bi#1 {\typeout{#1} \item}
\begin{document}
\title{Statistical mechanics of optimization problems}
\author{ Giorgio Parisi \\
  Dipartimento di Fisica, INFM, SMC and INFN, \\
Universit\`a di Roma {\em La Sapienza}, P. A. Moro 2, 00185 Rome, Italy. }
\maketitle
\begin{abstract}
Here I will present an introduction to the results that have been recently obtained in constraint
optimization of random problems using statistical mechanics techniques.  After presenting the
general results, in order to simplify the presentation I will describe in details the problems
related to the coloring of a random graph.

 \end{abstract}
 \section{Introduction}

In statistical mechanics \cite{I} the partition function is 
		  \be 
		  Z(\beta)=\sum_{\cC}\exp( -\beta H(\cC))\ ,
		  \ee 
		  where $\beta=1/(kT)$ and $\cC$ is a generic configuration.
			
In optimization problems \cite{II} we want to find the configuration $\cC^{*}$ that minimizes the
function $H(\cC)$ and to know the minimal cost, $ H^{*}\equiv H(\cC^{*}) $.  We define $E(T)$ and
$S(T)$ to be respectively the expectation value of the energy and the entropy as function of the
temperature $T$.  $ H^{*}$ is $E(0) $ and the number of optimizing configurations is $\exp(S(0))$.
In order to obtain information on the nearly optimizing configurations, i.e. those configurations
such that $H(\cC)=H^{*}+\eps$, we must study the system at small temperature.  We are interested to
knowing what happens in the thermodynamic limit, i.e. when the number $N$ of variables goes to
infinity.

In optimization theory it is natural to consider an ensemble of Hamiltonians and to find out the
properties of a generic Hamiltonian of this ensemble \cite{III,GaJo}, in the same way as in the 
theory of disordered systems.  Sometimes the ensemble is
defined in a loose way, e,g.  problems that arise from practical instance such as chips placements on
 computer boards. 
\Cred{In other words we have an Hamiltonian that is characterized by a set of parameters (denoted by $J$)
and we have a probability distribution $\mu(J)$ on this parameter space.}  We want compute the
ensemble average
		  \be
		  \int d\mu (J) E_{J}(T) \equiv \overline{E_{J}(T)} \ .
		  \ee

We are interested in computing the probability distribution $P(E)$ of the zero
temperature energy $E$ over the ensemble.
When the number $N$ of variables goes to infinity, if $E$ is well normalized, and its
probability distribution becomes a delta function \cite{sat0,KS,01}:
intensive quantity do not fluctuate in the thermodynamic limit.

\section{Constraint optimization}

In a typical case a configuration of our system is composed by $N$ variables $\sigma_{i}$ that 
take $q$ values (e.g. from 1 to $q$).  A instance of the problems is characterized by $M$ functions
$f_{k}[\sigma]$ ($k=1,M$), each function takes only the values 0 or 1.

		  Let us consider the following example with \Cblu{$N=4$ and $M=2$}:
		  \be
		  c_{1}[\sigma]=\theta(\si_{1}\si_{2}-\si_{3}\si_{4}) \ , \ \
		  c_{2}[\sigma]=\theta(\si_{1}\si_{3}-\si_{2}\si_{4}) \ .\nonumber  
		  \ee
		  The function we want to minimize is 
		  \be
		  H[\sigma]=c_{1}[\sigma]+c_{2}[\sigma] .
		  \ee

We are interested to know if there is a minimum with $H[\sigma]=0$.  If this happens all the
function $c_{k}$ must be zero.  The condition $H[\sigma]=0$ is equivalent to the following two
inequalities:
		  \be
		  \si_{3}\si_{4}>\si_{1}\si_{2} \ \ \ , 
		  \si_{2}\si_{4}>\si_{1}\si_{3} \ .
	%	  \si_{2}\si_{3}>\si_{1}\si_{4} \ .
		  \ee
Each function imposes a constraint: \Cred{the function $H$ is zero if all the
constraints are satisfied:}  we are in the
satisfiable case.  If not possible to satisfy all the constraints, the minimal
total energy is different from zero and we stay in the unsatisfiable case.

Given $N$ and $M$ we  define the ensemble as all the possible different sets of $M$ inequalities 
		  of the type
		  \be
		  \si_{i_{1}(k)}\si_{i_{2}(k)} >\si_{i_{3}(k)} \si_{i_{4}(k)}\ .
		  \ee
The interesting limit is when $N$ goes to infinity with 
		  \be
		  M=N\alpha \ , 
		  \ee		  
$\alpha$ being a parameter.  Hand waving arguments suggest that for small $\alpha$ it is should be
possible to satisfy all the constraints, while for very large $\alpha$  most of
the constraints will be not satisfied. 
We define the energy density
			  \be
		  e(\alpha)=\lim_{N\to\infty}\frac{\overline{H^{*}}}{N} \ .
		  \ee
There is a phase transition at a critical value of $\alpha_{c}$, such that
		  \be
		  e(\alpha)=0  \for \alpha<=\alpha_{c}\ , \ \ \ \
		  e(\alpha)>0  \for \alpha>\alpha_{c} \ .
		  \ee
	
\section{Random Graphs and Bethe approximation}

We define a random Poisson graph \cite{Erdos_Renyi} in the following way: given $N$ nodes we consider the
ensemble of \emph{all} possible graphs with $M=\alpha N$ edges (or links).  A random Poisson graph
is a generic element of this ensemble.			  
The local coordination number $z_{i}$ is  the number of nodes that are connected to the node $i$.
The average coordination number $z$ is given by $ z=2 \alpha $.
		
\Cred{These graphs are locally a tree}: if we take a generic point $i$, \Cblu{the subgraph composed
by those points that are at a distance less than $d$ on the graph is a tree} with probability one
when $N$ goes to infinity.  If $z>1$ the nodes percolate and a finite fraction of the graph belongs
to a single giant connected component.  Loops do exist on this graph, but they have typically a
length proportional to $\ln(N)$.  The absence of small loops is crucial because we can study the
problem locally on a tree and we have eventually to take care of the large loops as self-consistent
boundary conditions at infinity.

Random graphs are sometimes called Bethe lattices, \Cred{because a spin model on such a graph has the
moral duty to be soluble exactly using the Bethe approximation.}  Let us recall the Bethe
approximation for the two dimensional Ising model \cite{TAP,HH}.
In the standard mean field approximation, one arrives to the well known equation
		  \be
		  m=\tgh( \beta J z m) \ ,
		  \ee
where $z=4$ on a square lattice ($z=2d$ in $d$ dimensions) and $J$ is the spin coupling.  The
critical point is $\beta_{c} =1/z$.  This result is not very exciting in two dimensions (where
$\beta_{c}\approx .44$) and it is very bad in one dimensions (where $\beta_{c}=\infty$).  The aim of Bethe was to obtain a better results still keeping the simplicity of the mean field
theory.

\Cblu{Let us consider the system where a spin $\sigma$ has been removed}.  There is a cavity in the system
and the spins $\tau$ are on the border of this cavity.  \Cred{We assume that these spins are uncorrelated
and they have a magnetization $m_{C}$.}  When we add the spin $\sigma$, we find that the probability
distribution of this spin is proportional to
\be
\sum_{\tau_{i}}  \exp\left(\beta J \sigma \sum_{i=1,4}\tau_{i}\right) \prod_{i=1,4}P_{m_{C}}[\tau_{i}])\ .
\ee
The magnetization of the spin $\sigma$ is thus
\be
m=\tgh\{z \;\arth[ \tgh(\beta J) m_{C}] \}\ , \label{BETHE}
\ee
with $z=4$.

\Cblu{Now we  remove one of the spin $\tau_{i}$ and form a larger cavity (two spins removed).}
		  \Cred{We  assume that the spins on the border of the cavity are uncorrelated and they
		  have the same magnetization $m_{C}$.} We obtain
		  \be
		  m_{C}=\tgh\{(z-1) \arth[ \tgh(\beta J) m_{C}] \}\ . \label{CAVITY}
		  \ee
		  Solving this last equation we can find the value of $m_{C}$ and using the previous equation we can 
		  find the value of $m$.
			  It is rather satisfactory that in 1 dimensions ($z=2$) the cavity equations become
		  \be
		  m_{C}=\tgh(\beta J) m_{C}\ .
		  \ee
		  This equation for finite $\beta$ has no non-zero solutions, as it should be.
The internal energy and the free energy can also be  computed. The result is not exact because the 
cavity spins are correlated.

If we remove a
node of a random lattice \cite{MP1,MP2}, the nearby nodes (that were at distance 2 before) are now at a very large
distance, i.e. $O(\ln(N))$ with probability one.  In this case we hope that we can write
		  \be
		  \lan \tau_{i_{1}}\tau_{i_{2}} \ran \approx m_{i_{1}}m_{i_{2}}\ .
		  \ee
This happens in the ferromagnetic case  in presence if a in infinitesimal of magnetic field  where the
magnetization may take only one value.  
In more complex cases, (e.g. antiferromagnets)
\Cblu{there are many different possible values of the magnetization because there are many equilibrium
states. 
The cavity equations become equations
for the probability distribution of the magnetizations.}
This case have been long studied in the literature and we say that the replica symmetry is
spontaneously broken \cite{MPV,PBOOK}.  Fortunately for the aims of this talk we need only a very simple form of
replica symmetry breaking and we are not going to describe the general formalism.  

\via

\section{Coloring a graph}

For a given graph $G$ we would like to know if using $q$ colors the graph can be colored in such a way that
adjacent nodes have different colors \cite{MPWZ}.  The Hamiltonian is
		  \be
		  H_{G}=\sum_{i,k}A(i,k)\delta_{\sigma_{i},\sigma_{i}}\ ,
		  \ee
where $A_{G}(i,k)$ is the adjacency matrix and the variables $\sigma$ may take values that go from 1 to $q$.
\Cred{This Hamiltonian describes the antiferromagnetic Potts model with $q$ states.} For large $N$ on a random
graph energy density does not depend on $G$:
		  \be
		  e(z) =0 \for z<1\ , \ \ \
		  e(z) \propto \sqrt{z} \for z \to \infty \ .
		  \ee
There is a phase transition at $z_{c}$ between the colorable phase $e(z)=0$ and the uncolorable
phase $e(z)\ne 0$.  
For $q=2$  we have $z_{c}=1$: Odd loops cannot be colored and  for $z>1$ there are many large loops
that are even or odd with equal probability. 
The $q=2$ case is an antiferromagnetic Ising model on
a random graph, i.e. a standard spin glass.

Let us consider a legal coloring (i.e all adjacent nodes
have different colors).  We take a node $i$ and we consider the subgraph of nodes at distance $d$
from a given node.  Let us call $\cB(i,d)$ the interior of this graph.  
We ask the following questions: 
\begin{itemize}
	 \item Are there other legal colorings of the graph that coincide with the original coloring
	 outside $\cB(i,d)$ and differs inside $\cB(i,d)$?  We call the set of all these coloring
	 $\cC(i,d)$.
	\item
		  Which is the list of colors that the node $i$ may have in one of the coloring belonging to 
		  $\cC(i,d)$? We call this list $L(i,d)$. This list depends on the legal coloring we started from.
\end{itemize}

$L(i,d)$ has a limit when $d$ goes to infinity.  \Cred{We call this limit $L(i)$, i.e. the list of
all the possible colors that the site $i$ may have if we change only the colors of the nearby nodes
and we do not change the colors of faraway nodes.}

\Cblu{Let us study what happens on a graph where the site $i$ has been removed.} We denote by $k$ a
node adjacent to $i$ and we call $L(k;i)$ the list of the possible colors of the node $k$.
\Cred{The various nodes $k$ do not interact directly and their colors are independent.}

In this situation it is evident that $L(i)$ can be written as function of all the $L(k;i)$. 
We have
to consider all the neighbors ($k$) of the node $i$; if a neighbor may be colored in two ways, it
imposes no constraint, if it can be colored in only one way, it forbids the node $i$ to have its
color.  Considering all nearby nodes we construct the list of the forbidden colors and 
the allowed colors are those colors that are not forbidden.
	
A further simplification  may be obtained if we associate to a list $L$ a
variable $\omega$, that take values from 0 to $q$, defined as follow
\begin{itemize}
	 \item The variable $\omega$ is equal to $i$ if the list contains only the $i^{\mbox{th}}$ color.
	 \item The variable $\omega$ is equal to 0 if the list contains more than one color.
\end{itemize}

\Cblu{In the nutshell we have introduced an extra color, white. A site is white if it can
be colored in more than two ways without changing the colors of the far away sites \cite{P1,P5}.}
\Cred{The  equations are just the generalization of the Bethe equation where we have the colors,
white included, instead of the magnetizations.}  We have discrete, not continuos variables, because
we are interested in the ground state, not in the behavior at finite temperature. 
The previous equation are called the {\sl belief} equations or TAP equations.  We can associate to any
legal coloring a solution of the belief equations.  Sometimes the solution of the belief equations
is called a {\sl whitening}, because some nodes that where colored in the starting legal
configuration becomes white.

Each legal coloring has many other legal
colorings nearby that differs only by the change of the colors of a \textit{small} number of nodes.
\Cred{The number of these legal coloring that can be reached starting from a given coloring by making this
kind of moves is usually exponentially large and correspond to the same whitening.}
			
We have  three possibilities. 
\begin{itemize}
	 \item 
	 For all the legal configurations the corresponding whitenings have  all nodes  white.
	 \item
	 For a generic  legal configurations the corresponding whitening is non-trivial. 
	 i.e. for a finite fraction of the nodes are not white.
	 \item The graph is not colorable and there are no legal configurations.
\end{itemize}
			
In the second case we want to know the number of whitenings , how they differs and which
are their properties, e.g. how many sites are colored.  
In this case the set of all the legal
configurations breaks in an large number of different disconnected regions that are called with many
different names  (states, valleys, clusters,
lumps\ldots).  Each whitening is associated to a different
cluster of legal solutions \cite{P1}.
			
We consider the case where there is a large number of non-equivalent whitening We introduce the
probability $P_{i}(c)$ that for a generic whitening we have that $\omega(i)=c$.  The quantities
$P_{i}(c)$ generalize the physical concept of magnetization.
We will assume that for points $i$ and $l$ that are far away on the graph the probability
$P_{i,l}(c_{1},c_{2})$ factorizes into the product of two independent probabilities
		  $
		  P_{i,l}(c_{1},c_{2})=P_{i}(c_{1})P_{l}(c_{2}).
		  $	  
This hypothesis in not innocent: there are many cases where it is not correct.

A similar construction can be done with the cavity coloring and in this way we define the
probabilities $P_{i;k}(c)$, where $k$ is a neighbor of $i$.  These probabilities are called
{\bf surveys} \cite{MPZ,MZ,P3}.
Under the previous hypothesis the surveys satisfy  equations (the so called survey propagation
equations) that are simple, but are lengthy to be written.		
The survey propagation equations always have a trivial solution corresponding to all sites white:
$P_{i}(0)=1$ for all $i$.  Depending on the graph there can be also non-trivial solutions of the
survey equations.  Let us assume that if such a solution exist, it is unique.

We are near the end of our trip.  If we consider the whole graph we can define the probability $\cP[P]$, i.e.
the probability that a given node has a probability $P(c)$.  With some work one arrives to an integral
equation for $\cP[P]$, i.e. the probabilities of the surveys, whose solution can be easily
computed numerically  on present days computers.
One finds that there is a range $z_{d}<z<z_{U}$ where the previous integral equation has a
non-trivial solution and its properties can be computed.
			
In the same way that the entropy counts the number of legal colorings, the complexity counts the
number of different whitening; more precisely for a given graph we write
			\be
			\# \mbox{whitenings}= \exp (\Sigma_{G}) \ ,
			\ee
			where $\Sigma_{G}$ is the complexity.
			
\Cred{There is a simple way to compute the complexity.  It mimics the standard computation of the
free energy and it consists in counting the variation in the number of whitenings when we modify the
graph.} At the end of the day we find the results shown in fig.(1).

\begin{figure}
	 \begin{center}	 
		  \includegraphics[angle=0,width=0.5\columnwidth]{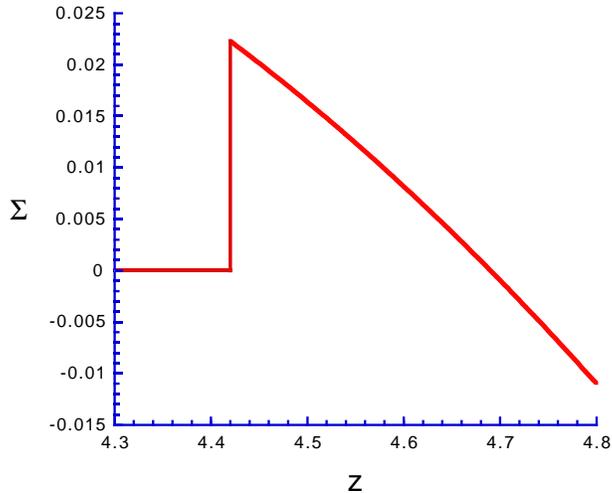}
		  \caption{The complexity versus the average connectivity $z$
	for three colors.}
\end{center}
 \end{figure}
he complexity jumps from 0 to a finite value at $z_{d}=4.42$; it decreases with increasing $z$
and becomes eventually negative at $z=4.69$.  A negative value of $\Sigma$ implies a number of
whitenings less than one and it is interpreted as the signal there there are no whitening (and no
legal configurations).  
\Cred{In the region
where the complexity is negative a correct computation of the energy $e(z)$ gives a non-zero
(positive) result.  The value where the complexity becomes zero is thus identified as the
colorability threshold $z_{c}=4.69$.} Similar results may be obtained for higher values of $q$ \cite{MPWZ}.
The concept of complexity emerged in
the study of the spin glasses and it was also introduced in the study of glasses under the name of
configurational entropy.  The behavior of the complexity as function of $z$ is very similar to what
is supposed to happen in glasses as function of $\beta$ \cite{PARISILH,LET}.

\section{Open problems}

There are many problems that are still open:
\begin{itemize}

	 \item The extension to other models.  
	 \item Verification of the self-consistency of the different hypothesis.  
	 \item The construction of effective algorithms for finding a solution of the optimization
	 problems.  A first algorithm has been proposed and it has been later improved by adding
	 backtracking \cite{P3,PB}.  A goal is to produce an algorithm that for large $N$ finds a solution on a random
	 graph in a polynomial time as soon as $z<z_{c}$.  Finding this algorithm is interesting from the
	 theoretical point of view (it is not clear at all if such an algorithm does exist) and it may
	 have practical applications.
	 \item  One should be able to transform the results derived in this way into rigorous theorems.
	 After a very long effort Talagrand \cite{Tala}, using some crucial
	 results of Guerra \cite{GUERRA}, has been recently able to prove that a similar, but more
	 complex, construction gives the correct results in the case of infinite range spin glasses, i.e.
	 the Sherrington Kirkpatrick model, that was the starting point of the whole approach. Some of 
	 these results have been extended to the case of the Bethe Lattice \cite{FraLeo}.

\end{itemize}


\begin{thebibliography}{100}

\bibitem {I} See for example: Parisi G. {Statistical Field Theory} (Academic Press, New York) 1987.
	 
\bibitem{II} Martin O. C., Monasson R. and Zecchina R., \emph{Theoretical Computer Science}
\textbf{265} (2001) 2.

\bibitem{III} G.Parisi \emph{Constraint Optimization and Statistical Mechanics}, cond-mat/0301157 (2003).

\bibitem{GaJo} Garey M. R. and Johnson D. S., \textit{Computers and intractability} (Freeman, New
York) 1979.

\bibitem {sat0}Dubois O. Monasson R., Selman B. and Zecchina R., \emph{Phase Transitions in
Combinatorial Problems}, \textit{Theoret.  Comp.  Sci.} \textbf{265}, (2001), G. Biroli, S. Cocco, R.
Monasson, \textit{Physica A} \textbf{306}, (2002) 381.

\bibitem{KS}  Kirkpatrick S. and  Selman B., \emph{Critical Behaviour in the
satisfiability of random Boolean expressions}, \textit{Science} \textbf{264}, (1994) 1297.

\bibitem{01} Dubois  O., Boufkhad Y., Mandler J., \emph{Typical random 3-SAT formulae and the
satisfiability threshold}, in {\it Proc.  11th ACM-SIAM Symp.  on Discrete Algorithms}.

	
	 \bibitem{Erdos_Renyi} P. Erd\"os and A. R\`enyi, Publ. Math. 
	 (Debrecen) {\bf 6}, 290 (1959).

	
	 \bibitem{TAP} Thouless D.J., Anderson P.A. and Palmer R. G., \emph{Phil.  Mag.} \textbf{35}, (1977)
	 593.
	 \bibitem{HH}Katsura S., Inawashiro S. and Fujiki S., \emph{Physica} \textbf{99A} (1979) 193.

	


\bibitem{MP1}   M\'ezard M. and  Parisi G..  \emph{Eur.Phys. J.} B {\bf 20} (2001) 217.

\bibitem{MP2}  M\'ezard M. and  Parisi G..	 \textit{J. Stat. Phys} {\bf 111}, (2003) 1 .

\bibitem{MPV} M\'ezard, M., Parisi, G. and Virasoro, M.A.\emph{ Spin Glass Theory and Beyond}, (World
Scientific, Singapore) 1997.

\bibitem{PBOOK} Parisi G., {\sl Field Theory, Disorder and Simulations}, (World Scientific,
Singapore) 1992.

	 
\bibitem{MPWZ} Mulet R., Pagnani A., Weigt M., Zecchina R., {\it Phys.  Rev.  Lett.} {\bf 89}, 268701
(2002); Braunstein A., Mulet R., Pagnani A., Weigt M., Zecchina R., \textit{Phys.  Rev.  E}
\textbf{68}, (2003) 036702.
	

\bibitem{P1}Parisi G.. cs.CC/0212047 \emph{On local equilibrium equations for clustering states}
(2002).

\bibitem{P5} Parisi G., \textit{On the probabilistic approach to the random satisfiability problem}
 cs.CC/0308010 (2003).



\bibitem {MPZ} M{\'e}zard M., Parisi G. and Zecchina R., \emph{Science} \textbf{297}, (2002) 812.
	
	 \bibitem {MZ}  M{\'e}zard  M.and  Zecchina  R. {\em Phys. Rev. E} {\bf 66}, 056126 (2002).

	 \bibitem{PARISILH}  Parisi G., \emph{Glasses, replicas and all that} cond-mat/0301157 (2003).

		 \bibitem{LET}  Cugliandolo T.F., \emph{Dynamics of glassy systems } cond-mat/0210312  (2002).

	 \bibitem{P2} Parisi. G.  \emph{On the survey-propagation equations for the random
	 K-satisfiability problem} cs.CC/0212009 (2002).

	 
	 \bibitem{P3} Parisi. G.  \emph{Some remarks on the survey decimation algorithm for
	 K-satisfiability} cs.CC/0301015 (2003).

		 
	 \bibitem{PB} Parisi G. \emph{A backtracking survey propagation algorithm for K-satisfiability},
	 cond-mat/0308510 (2003).

	 \bibitem{Tala} Talagrand M. {\em Spin Glasses. A challenge for mathematicians.  Mean-field models
	 and cavity method}, (Springer-Verlag Berlin) 2003,  \emph{The Parisi formula}. 
% 	 Annals of Mathematics, to appear.
	 
	 \bibitem{GUERRA} Guerra F., \emph{Comm.  Math. Phys.} {\bf 233} (2002) 1; Guerra F. and 
	 Toninelli F.L., \emph{Comm.  Math. Phys.} {\bf 230} (2002), 71.
	 
	 \bibitem{FraLeo} S. Franz and M. Leone, \emph{J. Stat. Phys} \textbf{111} (2003) 535. 


\end{thebibliography}
\end{document}